\documentclass[aps,pra,nofootinbib]{revtex4}
\usepackage{amsbsy,latexsym}
\usepackage{amsfonts}
\usepackage{amssymb}
\usepackage{graphicx}
\usepackage[mathscr]{eucal}
\newcommand{\eqref}[1]{(\ref{#1})}

\newcommand{\carac}[1]{\chi_{#1}}
\newcommand{\al}{\alpha}
\newcommand{\be}{\beta}
\newcommand{\ga}{\gamma}
\newcommand{\de}{\delta}
\newcommand{\la}{\lambda}
\newcommand{\id}{\mathbb I}
\newcommand{\CF}{\mathsf n}
\newcommand{\aver}[1]{\langle #1 \rangle}

\newcommand{\ket}[1]{\vert #1 \rangle}
\newcommand{\bra}[1]{\langle #1 \vert}

\newcommand{\D}[3]{\mathfrak D^{(#1)}_{#2 #3}}

\newcommand{\Chi}[1]{\chi^{(#1)}}

\begin{document}
\title{ALIGNING SPATIAL FRAMES THROUGH QUANTUM CHANNELS}
\author{E.~Bagan \& R.~Mu{\~n}oz-Tapia}
\address{Grup de F{\'\i}sica Te{\`o}rica \& IFAE,
Facultat de Ci{\`e}ncies,  Edifici Cn, \\
Universitat Aut{\`o}noma de Barcelona, 08193
Bellaterra (Barcelona) Spain}

\begin{abstract}

We review the optimal protocols for aligning spatial frames using
quantum systems. The communication problem addressed here concerns
a type of  information that cannot be digitalized. Asher Peres
referred to it as ``unspeakable information". We comment on his
contribution to this subject and give a brief account of his
scientific interaction with the authors.

\end{abstract}

\maketitle
\section{Introduction}
In his later years, Asher Peres became interested in a type of
information that he suggestively called ``indiscrete" and, later
on, ``unspeakable" ---paraphrasing the title of J.~S.~Bell's
famous book~\cite{bell}. In Asher's own words~\cite{unspeakable}:
\begin{quote}{\em
The role of a dictionary is to
define unknown words by means of known ones. However there are
terms, like left or right, which cannot be explained in this way.
In the absence of a formal definition, material objects must be
used to illustrate these terms: for example, we may say that the
human liver is on the right side. Likewise, the sign of helicity
may be referred to the DNA structure, or to the properties of weak
interactions.}
\end{quote}
In this context, there are interesting questions which can be
readily translated into quantum information problems. Imagine Bob
went on an intergalactic journey and because of a classical
computer failure, he is now completely lost in space. Alice and
Bob can  communicate through a classical channel, but there is no
way for them to establish a common reference frame. To help Bob
come back home, Alice needs to send
him information about vectors in space (about the one that points home). 
In these circumstances sending any string of bits has no meaning
whatsoever, since there is no shared reference frame to which Bob
can refer them. Alice can only send Bob information
``analogically":  e.g., she can send a physical object that
carries intrinsic orientation. In the quantum world there are such
kind of objects: particles with spin. So, the question is: how
efficiently can one communicate directions, align reference
frames, etc. using such particles?

About five years ago, we became interested in this kind of topic.
Sandu Popescu, who at the time visited Barcelona, explained to us
that a state of two antiparallel spin-1/2 particles (spins for
short) codifies space directions (three dimensional unit vectors)
better than a state of two parallel spins~\cite{gisin-popescu}.
Then the obvious questions arose:  what is the best quantum state
for this purpose if $N$ spins are available? what is the best
measurement to retrieve this information?

Shortly after we published  the answer to these questions and the
general solution to the $N$-copy problem~\cite{bbm-direction}, a
paper by Asher and  Petra Scudo, who was Asher's student at the
time, appeared in the archive~\cite{ps-direction}.
The paper addressed the same problem from a more physical point of
view. They were not that interested in analytical  results, and
rather resorted to clever numerical analysis. Minor disagreements,
which were finally overcome, triggered a very intense e-mail
exchange out of which we got the chance to get acquainted with
Asher.

Having raised the question of how to communicate through a quantum
channel when the parties share no reference frame, the natural
next question we wished to address was how to optimally align
spatial frames (SF) ---i.e. orthogonal trihedra--- using this
channel.
A few months later our analysis was completed, and before we could
even start writing  a draft a new paper by Asher and Petra
appeared in the archive~\cite{ps-reference}. As it turned out,
once again we had had the same agenda, tough this time Asher and
Petra were quicker than we were. Their numerics revealed some
stronger disagreement with our results and `there we go again!' We
exchanged e-mails for weeks, trying to figure out what was wrong,
until we all agreed. In the meantime, we completed our draft and
got it published~\cite{bbm-single} some months later. We got a lot
out of this interaction, because Asher's physical insight was
astonishingly deep and he was always eager to share, but a great
deal  of it goes far beyond the scientific realm.

Our approaches to the alignment of SFs were slightly different.
While Asher and Petra considered superpositions of orbital angular
momentum states ---such as those of a hydrogen atom---  we focused
on systems made out of $N$ spins. The two approaches are formally
equivalent if $N$ is even and if in the latter the multiplicity of
the irreducible representations that show up in the Clebsch-Gordan
decomposition of $({\bf 1/2})^{\otimes N}$ is not taken into
account. The two approaches boil down to a sort of rigid covariant
encoding of $SO(3)$ rotations into a quantum state of spins using
that $SO(3)\simeq SU(2)$. As we see it now, covariance is inherent
to the problem at hand. In the absence of a shared frame, all
Alice can do is rotate (a fixed anisotropic state of) a system and
send it to Bob, who can afterwards attempt to align his spacecraft
(his SF) with the system.

Shortly after, yet another new paper~\cite{ps-noncovariant} by
Asher and Petra challenged us.
They pointed out that by splitting the system of spins into three
sets, and using them to transmit the directions of each of the
three axis of the SF separately, one could outstrip the
performance of our frame-transmission protocol.
Actually, by that time, we had realized that our protocol was less
efficient than that Asher was then proposing, and we were already
looking for improvements. Asher and Petra's paper boosted our
commitment to the search we had already undertaken, but it took us
quite a while to complete~\cite{bbm-entangled,bbm-repeated}. In
the meantime Asher, together with Netanel Lindner and Daniel
Terno, considered a concrete realization of  the alignment
problem.  They proposed to use Elliptic Rydberg
states~\cite{lindner}, which are the quantum analogues for the
hydrogen atom of the classical Keplerian orbits and are feasible
in a laboratory.

All this progress has triggered other interesting developments
such as e.g., the optimal protocols for communication of digital
information without shared reference
frame~\cite{bartlett,cabello}, an issue not far removed from
superselection rules~\cite{verstraete} or from the quantification
of the information contained in the establishment of a shared
reference frame~\cite{collins,enk}.

 This paper is a review in perspective of our present knowledge of
 optimal alignment of SFs using systems of spins.
 The paper is organized as follows. In the next section
 we formulate the problem and introduce the notation.
 In Sec.~\ref{sec:single} we discuss
 the optimal protocol for a hydrogen atom.
In Sec.~\ref{sec:repeated} we show that an efficient use of a $N$
spin system decreases the communication error drastically.
 In Sec.~\ref{sec:entangled} we present the absolute optimal
 protocol.
This protocol, which requires full-fleshed entanglement, can be
viewed as dense covariant coding of ``unspeakable information"
(using Asher's terminology) or continuous
variables~\cite{braunstein}. The last section contains our
conclusions.

\section{Preliminaries and Notation}\label{sec:notation}

To lend a dramatic touch to the presentation, let us consider he
situation described in the introduction, in which Bob's spacecraft
is lost in space while Alice, on Earth, is sending him (through a
quantum channel) ``unspeakable" information consisting of a SF
$\CF=\{\vec n_{1}, \vec n_{2},\vec n_{3}\}$. Let us assume that
Alice has a quantum system that she can prepare in a superposition
of states with angular momentum $j=0,1,2,\dots$, and let $J$ stand
for the maximum angular momentum in the superposition state. Prime
examples of such system are a hydrogen atom and a system of $N$
(even) spins.

By performing quantum measurements on the state that Alice has
sent to him, Bob can, in principle, estimate $\CF$ with some
accuracy. From the outcomes of his measurements he makes a guess
$\CF'=\{\vec n'_{1}, \vec n'_{2},\vec n'_{3}\}$ of the SF $\CF$. A
convenient parametrization of $\CF$ and $\CF'$ is given by the
Euler angles $\al$, $\be$, $\ga$ of the rotations that take an
external observer's ---e.g., the author's--- SF, $\CF_{0}=\{\vec
x,\vec y,\vec z\}$, into $\CF$ and $\CF'$.  We will use $g$ as a
shorthand for the three Euler angles, i.e., $g=(\al,\be,\ga)$. Of
course, neither Alice nor Bob need to be aware of the existence of
$\CF_{0}$. Following Holevo~\cite{holevo}, we use the error,
defined by
\begin{equation}
    h(g,g')=\sum_{a=1}^3|\vec n_{a}-\vec n_{a}'|^2=
    \sum_{a=1}^3|\vec n_{a}(g)-\vec n_{a}(g')|^2,
    \label{ebc-6.5-6}
\end{equation}
to quantify the quality of Bob's guess. Alice encodes $\CF$
[$=\CF(g)$] into a suitable quantum state $\ket{A(g)}$ of her
system. Covariance allows us to write
\begin{equation}\label{Ag}
    \ket{A(g)}=U(g)\ket{A},
\end{equation}
where $\ket{A}$ is a fixed fiducial state and $U(g)$ is the
unitary representation on Alice's (system) Hilbert space of the
rotation that takes $\CF_{0}$ into $\CF$.

Since Bob is completely lost in space, he has absolutely no
information about~$\CF$. Therefore,  if we wish to compute how
well Bob is doing on average, we must assume that {\em \`a priori}
Alice's SF are isotropically distributed, i.e.,  we must
define~$dg$ by the Haar measure of $SO(3)$~\cite{edmonds}, which
in terms of the Euler angles reads $dg=\sin\be\, d\be d\al d\ga
/(8\pi^2)$.

To compute the average error $\aver{h}$, the optimal measurement
---represented in full generality by a Positive Operator Valued
Measure (POVM)--- can be chosen to be covariant~\cite{holevo} and
to have a rank-one seed~\cite{davies}:
\begin{equation}\label{povm-single}
   O(g')=U(g')\, \ket{B} \bra{B}\,U^\dagger(g').
\end{equation}
The state $\ket{B}$ defines Bob's POVM much in the same way as
$\ket{A}$ defines Alice's messenger state.

The conditional probability of Bob guessing $\CF(g')$ if Alice's
SF is $\CF(g)$ is given by quantum mechanics through the Born rule
$p(g'|g)=\bra{A(g)} O(g') \ket{A(g)}$. Then the average error
reads
\begin{equation}
    \aver{h}\equiv\int dg\int dg' h(g,g') p(g'|g)=\int dg\, h({\bf 0},g) |\bra{A} U(g) \ket{B}|^2 ,
    \label{aver-h-2}
\end{equation}
where $\bf0$ stands for $(\al,\be,\ga)=(0,0,0)$. To derive the
second equality we have used the invariance of the Haar measure
and the normalization condition $\int dg=1$. One can easily check
that
\begin{equation}
    h({\bf0},g)=6-2\Chi1(g),
    \label{h-t}
\end{equation}
where, following a widely used notation, we denote by $\Chi j$ the
character of the spin-$j$ representation of $SO(3)$. Namely,
\begin{equation}\label{t}
\Chi j(g)\equiv\sum_{m} \D{j}{m}{m}(g) ,
\end{equation}
where the elements of the matrix $\D j m {m'}$ are defined in the
standard way as $\D{j}{m}{m'}(g)\equiv\bra{jm}U(g)\ket{jm'}$. One
also has
$\Chi1(g)=\cos\be+(1+\cos\be)\cos(\al+\ga)$, from which it follows
that the values of $\Chi1$ lie in the real interval $[-1,3]$. The
value $\Chi1=3$ corresponds to perfect determination of Alice's SF
and implies that $h=0$. Note also that $\aver{h}=6-2\aver{\Chi1}$.
Random guessing implies $\aver{\Chi1}=0$ ($\aver{h}=0$), while
perfect determination of one axis and random guessing of the
remaining two yield $\aver{\Chi1}=1$ ($\aver{h}=4$).

There is no consensus on the loss function for this
frame-transmission problem. Instead of $h$, Asher and
Petra~\cite{ps-reference} used the Mean Square Error per axis,
defined as $\mathrm{MSE}=1/6 \sum_a (1-\vec{n}_a\cdot
\vec{n}'_a)$. One can readily see that
$\mathrm{MSE}=(3-\aver{\Chi1})/6$. In the following, we will focus
on $\aver{\Chi1})$. From our results, values for the reader's
favorite loss function can be obtained trivially.

In the following sections we will show that perfect SF
transmission (i.e $\aver{\Chi1}\to 3$) is possible in the
asymptotic limit of $J\to\infty$ (or equivalently $N\to \infty$),
provided the appropriate  signal state $\ket A$ and POVM seed
state $\ket B$ are chosen. We will also compute the rate at which
this limit is approached for each of the protocols discussed
below.

\section{Hydrogen atom}\label{sec:single}
Following Asher's approach, consider that Alice prepares a
hydrogen atom in the state
\begin{equation}
    \ket{A}=\sum_{j} \ket{A^j}=\sum_{jm} A_{m}^j\ket{jm};\qquad
\sum_{jm}|A_{m}^j|^2=1.
    \label{A-single}
\end{equation}
A system of an even number $N$ of spins can also be prepared in
such a state if one neglects all but one of the equivalent
spin-$j$ representations that appear with multiplicity
\begin{equation}
n_j=
\begin{pmatrix}
N\\ N/2+j
\end{pmatrix}  {2j+1\over N/2+j+1}  \label{n_j}
\end{equation}
 in the Clebsch-Gordan
decomposition of $({\bf 1/2})^{\otimes N}$.
In~Eq.~\eqref{A-single}, $j$ runs from $0$ to $J$ ($N/2$), and $m$
runs from $-j$ to $j$. The state $\ket A$ is a $n$'th energy level
of a hydrogen atom (a Rydberg state)~\cite{ps-reference}. As
mentioned above, a single transmission of  the rotated state
$\ket{A(g)}$ in Eq.~\eqref{Ag} can give Bob (``unspeakable'')
information about Alice's SF. The larger  the value of $J$ (equiv.
$N$) the better Bob's guess will be and, as we will show,  perfect
determination is possible  in the asymptotic limit.

In full analogy with Eq.~(\eqref{A-single} we write
\begin{equation}
    \ket{B}=\sum_{j}\sqrt{d_j}\ket{B^j};\quad
    \ket{B^j}=\sum_{m} B_{m}^j\ket{jm},
    \label{ketB}
\end{equation}
where the square root of the dimension of the spin-j
representation, $d_j=2j+1$, is introduced for later convenience.
The condition $\id=\int dg\,O(g)$ requires that
\begin{equation}
    \sum_{m} |B_{m}^j|^2=1,\quad \forall j,
    \label{povm-condition}
\end{equation}
as follows from Schur lemma.

{}From Eq.~\eqref{aver-h-2} we have
\begin{equation}
    \aver{\Chi1}=\int dg\,|\bra{A}U(g)\ket{B}|^2\,\Chi1(g),
    \label{aver-t}
\end{equation}
and to optimize the transmission we just have to maximize
$\aver{\Chi1}$ over $A^j_m$ and $B^j_m$. Hence, we write
$\aver{\Chi1}_{\rm max}=\max_{AB}\aver{\Chi1}$. Rather than
attempting a numerical maximization, as Asher and Petra
did~\cite{ps-reference}, we chose to play with  some group theory
and a few algebraic tricks in order to find general analytical
solution to the problem.~\cite{bbm-single}. For this purpose, it
is convenient to rewrite Eq.~\eqref{aver-t} as
\begin{equation}
    \aver{\Chi1}=\sum_{lj}{\sqrt{d_l d_j}\over3}
    \bra{B^j\tilde B^l}P_{1}\ket{A^j\tilde A^l}.
    \label{aver-t-2}
\end{equation}
In this equation $\ket{A^j\tilde A^l}=\ket{A^j}\otimes\ket{\tilde
A^l}$, where the state $\ket{\tilde A^j}$ is the time reversed of
$\ket{A^j}$ [i.e., $\tilde A^j_{m}=(-1)^m A^{j*}_{-m}$ and
similarly for $\ket{B^j\tilde B^l}$ and $\ket{\tilde B^l}$] and
$P_{1}$ is the projector over the Hilbert space of the
representation of spin $j=1$. With the help of the  Schwarz
inequality, we find that the  maximum in Eq.~\eqref{aver-t-2}
occurs for states $\ket{A}$  such that
\begin{equation}
    A^j_{m}=C^j B^j_{m},\qquad\mbox{with}\quad \sum_{j}|C^j|^2=1.
    \label{A-prop-B}
\end{equation}
This equation just tells us that, for an optimal communication,
the messenger states $\ket{A(g)}$ must be as similar as possible
to the states $U(g)\ket{B}$ on which the measuring device
projects~\cite{bbm-direction}. We then have
\begin{equation}
    \aver{\Chi1}_{\rm max}=\max_{BC}\sum_{jj'}C^j\, {\mathsf M}^{jj'}_{B}\,
    C^{j'},
    \label{aver-t-3}
\end{equation}
where
\begin{equation}
    {\mathsf M}^{jj'}_{B}={\sqrt{d_j d_{j'}}\over3}
    \bra{B^j\tilde B^{j'}}P_{1}\ket{B^j\tilde B^{j'}} ,
    \label{matrix-1}
\end{equation}
and the maximization is over all $B^j_{m}$ and $C^j$ subject to
the normalizations~\eqref{povm-condition} and~\eqref{A-prop-B}.
The maximum $\aver{\Chi1}_{\rm max}$ is thus given by the largest
value, $\lambda_{\rm op}$, in the set $\{\lambda_B\}$ of largest
eigenvalues of matrices of the form~\eqref{matrix-1}. Notice that
the entries of all these matrices are non-negative. In this
situation, it is easy to see that if a matrix $\mathsf{M}_{\rm
op}$, such that $\mathsf{M}_{\rm op}^{ij}\geq \mathsf{M}^{ij}_B$
for all $B$,  exists, then $\lambda_{\rm op}$  is precisely its
largest eigenvalue. Notice also that all the matrices
in~\eqref{matrix-1} are tri-diagonal. It is easy to
verify~\cite{bbm-single} that such ${\mathsf M}_{\rm op}$ indeed
exists and is given by
\begin{equation}\label{matrix-old}
\mathsf{M_{\rm op}}=\pmatrix {{J\over J+1}&\sqrt{2J-1\over2J+1}&
         &        &       \cr
                    \sqrt{2J-1\over2J+1}&\ddots&\ddots
    &\phantom{\ddots} \raisebox{2.0ex}[1.5ex][0ex]{\LARGE
0}\hspace{-.5cm}       &       \cr
                         &\ddots&{2\over3}    &\sqrt{{3\over5}} &       \cr
                         & \phantom{\ddots}
                         &\sqrt{{3\over5}}&{1\over2}&\sqrt{{1\over3}}
\cr \hspace{.5cm} \raisebox{2.0ex}[1.5ex][0ex]{\LARGE
0}\hspace{-.5cm} &&\phantom{\ddots}&\sqrt{{1\over3}}&0 } .
\end{equation}

A straightforward choice of the state $\ket{B}$ that saturates
these bounds is given by
\begin{equation}
    \ket{B_{\rm op}}=\sum_{j}\sqrt{d_j}\ket{j,j}\quad\Leftrightarrow\quad
    B^j_{{\rm op}\,m}=\de^j_{m}.
    \label{B-optimal}
\end{equation}
The form of the optimal seed state $\ket{B_{\rm op}}$, which in
turn determines through Eq.~\eqref{A-prop-B} Alice's optimal
messenger state,  agrees with our physical intuition. If Alice's
state had a well defined total spin along some axis (i.e., if it
were an eigenstate of $\vec n\cdot\vec J$, for some unit vector
$\vec n$), $\mathsf{M}_{\rm op}$ would become diagonal and
$\aver{\Chi1}_{\rm max}=J/(J+1)=N/(N+2)$ thus, at most (in the
limit $N\to\infty$) $\aver{\Chi1}=1$. In average, Bob could not
determine more than just one axis of Alice's SF. The structure of
the state $\ket{B_{\rm op}}$ is  such that,  within each
irreducible representation, the determination of a single axis is
optimal~\cite{bbm-direction-2} (this is {\em the best} Alice could
do if she only were allowed to use a single irreducible
representation). At the same time, $\ket{B_{\rm op}}$ is as
different from an eigenstate of $\vec n\cdot\vec J$ as it can
possibly be.

The problem is now solved; the optimal measurement is determined
by Eq.~\eqref{B-optimal},  $\aver{\Chi1}_{\rm max}$ is given by
the largest eigenvalue of $\mathsf{M}_{\rm op}$, and the optimal
messenger state, $\ket{A}$, is computed through
Eq.~\eqref{A-prop-B} substituting $C^j$ by the corresponding
eigenvector $C^j_{\rm{op}}$. For small $N$, one can easily obtain
analytic expressions for $\aver{\Chi1}_{\rm max}$, $\ket B$
and~$\ket{A}$.

In the asymptotic regime of large $J$ ($N$), it is again possible
to compute $\aver{\Chi1}$ explicitly up to sub-leading order in
$1/J$ ($1/N$). Sub-leading orders are important because, among
other things,  they enable us to compare different acceptable
protocols (those that achieve perfect determination in the strict
limit $J, N\to\infty$) independently of $J$ ($N$), as these orders
tell us the rate at which perfect determination
($\aver{\Chi1}_{\rm max}\to 3$) is reached. They are also very
important in quantum statistics~\cite{masahito}. For that purpose
we give simple upper and lower bounds of $\aver{\Chi1}_{\rm max}$.
 A useful upper bound is provided
by the condition
\begin{equation}
\aver{\Chi1}_{\rm max}\leq \max_{j}\sum_{j'}
    {\mathsf M}_{\rm op}^{jj'} ,
\end{equation}
    while a lower bound can be derived
    (with hard work and determination)
  using a variational method with a judicious choice of the
    vector $C^j$ (see Ref.~\cite{bbm-single} for details). We obtain
\begin{equation}
  3-{4 \over N}+O(N^{-4/3})\leq \aver{\Chi1}_{\rm max}\leq 3-{4\over
    N}+O(N^{-2}).
    \label{aver-t-asymptotic}
\end{equation}
We see  that perfect determination of Alice's SF is attained at a
rate linear in~$1/N$.

\section{$N$ spins. Use of the equivalent representations.}\label{sec:repeated}

We will now show that with $N$ spins Alice can devise a
communication protocol that outperforms that of the previous
section, provided she makes proper use of some of the $n_j$
equivalent spin-$j$ representations of $SO(3)$.

Instead of~\eqref{A-single}, the most general state in which Alice
can actually prepare her $N$-spin system is
\begin{equation}
|A\rangle=\sum_j\ket{A^j}=\sum_{j m \alpha}
A^j_{m\alpha}|jm\alpha\rangle, \qquad \sum_{j m \alpha}
|A_{m\alpha}^j|^2=1, \label{general-A}
\end{equation}
where we have introduced the additional index $\al$ to label the
$n_j$ equivalent
spin-$j$ representations. Recall that the index $\al$ does not rotate under $SO(3)$ 
and that the corresponding irreducible spaces are orthogonal,
namely,
$\bra{jm\al}U(g)\ket{jm'\al'}=\de_{\al\al'}\D{j}{m}{m'}(g)$.
This index $\al$ corresponds to a truly additional degree of
freedom of the system and, based on the work by Ac\'{\i}n {\em et
al.}~\cite{ajv}, one could argue that by entangling it with the
magnetic number $m$ [which does rotate under $SO(3)$] one could
improve on the previous protocol (see also next section).
\begin{equation}
\ket{A^j}={1\over\sqrt{d_j}}\sum_{m}\ket{jm\alpha_m}. \label{Aj}
\end{equation}
Note that  this (maximum) entanglement of degrees of freedom can
be established on any of the spin-$j$ invariant subspaces but on
the $j= J$ one ($J\equiv N/2$ throughout this section), which
corresponds to the highest spin. This is so because from
Eq.~\eqref{n_j} one has $n_{J}=1$, whereas $ n_j\ge d_j$ if $j<J$.
Therefore, $\ket{A^{J}}=\sum_m A^{J}_m\ket{Jm}$ has no
entanglement of this type at all.

It is not difficult to convince oneself that the $n_j-d_j$
equivalent representations that do not show up in Eq.~\eqref{Aj}
are actually sterile, i.e we cannot make them play any role in the
estimation problem at hand.
Recalling from Refs.~\cite{bbm-single} and~\cite{holevo} that
$\ket{JJ}$ is optimal when only one of the equivalent
representations is allowed, we propose the following fiducial
messenger state
\begin{equation}
\ket{A}=a_J \ket{JJ}+\sum_{j<J}
{a_j\over\sqrt{d_j}}\sum_{m}\ket{jm\alpha_m} \label{new-A}
\end{equation}

A covariant  POVM for these signal states is given by
Eq.~\eqref{povm-single},
where $\ket{B}$ is now chosen to be
\begin{equation}
   \ket{B}=\sum_{j}
\sqrt{d_j}\sum_{m}\ket{jm\alpha_m}. \label{new-B}
\end{equation}
%
Substituting \eqref{new-A} and \eqref{new-B} in \eqref{aver-t} we
can write $\aver{\Chi1}$ as the quadratic form [which plays the
role of~\eqref{aver-t-3}]
\begin{equation}\label{qform}
\aver{\Chi1}_{\rm max} = \max_{\mathsf a} \;\mathsf{a}^t(1+
\mathsf{M}) \mathsf{a},
\end{equation}
where ${\mathsf a}^t=(a_J,a_{J-1},a_{J-2},\dots)$ in the transposed of $\mathsf a$.
 Considering for
simplicity $N$ odd (i.e. $J$ half integer or $N=2n+1$), the matrix
$\mathsf{M}$ is
\def\xxx{\phantom{\displaystyle{.\over .}}}
\def\zzz{\hspace{.5em} }
\begin{equation}\label{matrix}
\mathsf{M}=\pmatrix{
 {-1\over J+1}     & \zzz{1\over \sqrt d_J}              &         &   &  &  &\xxx\cr
 {1\over \sqrt d_J}&\zzz 0                               & \zzz  1     &   & & \mbox{\LARGE0} &\xxx\cr
                   & \zzz1 & \zzz 0 &\zzz 1                   &         &   &\xxx\cr
                   &   &\zzz\raisebox{1.3ex}[0ex][0ex]{1}&\zzz\zzz  \ddots &\zzz \ddots & &\xxx\cr
                   &   &                             &\zzz \ddots  &       &\zzz \zzz1&\xxx\cr
 \hspace{.5cm} \raisebox{2.0ex}[1.5ex][0ex] {\LARGE0}\hspace{-.5cm} & & & &\zzz\zzz  1  &\zzz\zzz 0 &\zzz \zzz1\xxx\cr
                   &   &   &  &  & \zzz\zzz   1  &   \zzz  \zzz    0\xxx
                   } \ ,
\end{equation}
 The maximum value of \eqref{qform} is
$\aver{\Chi1}=1-2\lambda_0$, where $-2\lambda_0$ is the largest
eigenvalue of $\mathsf M$. The characteristic polynomial of
$\mathsf M$, defined here  to be $P^{J}_n(\lambda)=\det({\mathsf
M}+2\lambda)$, satisfies the recursive relation of the Tchebychev
polynomials~\cite{abramowitz}, hence, $P^{J}_n(\la)$ is a linear
combination of them. One can check that the explicit solution is
\begin{equation}
P_n(\la)=U_n(\la)-{2\over2n+1}
U_{n-1}(\la)+{2n-1\over2n}U_{n-2}(\la),
\end{equation}
where $P_n(\lambda)\equiv P^{n-1/2}_n(\la)$ and $U_n$ are the
Tchebychev polynomials
\begin{equation}
U_n(\cos\theta)=\sin[(n+1)\theta]/\sin\theta .
\end{equation}
Hence, the smallest zero of $P_n(\la)$, which we write as
$\la_0\equiv\cos\theta_0$, can be easily computed in the large $n$
limit expanding around $\lambda_0=-1$, i.e,
$\theta_0=\pi(1-n^{-1}+a n^{-2}+b n^{-3}+\dots)$. We find
\begin{equation} \label{avert-t-5}
\aver{\Chi1}_{\rm max}=3-{4\pi^2\over N^2}+{8\pi^2\over N^3}+\dots
.
\end{equation}
(See Refs.~\cite{giulio} and~\cite{masahito-frame} for alternative
derivations of this equation.) By comparing with
Eq.~\eqref{aver-t-asymptotic}, we note the communication protocol
presented here outperforms that of the previous section, as
already stated above. The rate at which it achieves perfect
determination of Alice's SF is quadratic in $1/N$, in contrast to
the former protocol which attains this limit linearly in $1/N$.
Despite this success, from our derivation it should be clear to
the reader  that this protocol may not be optimal. For finite $N$
this is indeed the case. However, the results in the next section
show that the protocol is \emph{asymptotically} optimal.

\section{Optimal protocol. Dense covariant coding.}\label{sec:entangled}

The optimal protocol requires that Alice and Bob share a maximally
entangled state. It bears a great similarity with dense
coding~\cite{dense} as far as the use it makes of entanglement.
However, the information we are attempting to transmit has an
``unspeakable" nature. Our derivation partially rely on the
results of Ref.~\cite{ajv}, where the optimal (entangled) state
for encoding an $SU(2)$ operation was obtained.

We now assume that Alice and Bob have {\em each of them} a system
of $N$ spins. Before Bob's departure for the intergalactic
journey, they prepare a suitable entangled state $\ket{\Phi}$.
Also before departure, Bob locks the orientation of his quantum
system to that of his spacecraft (SF and spacecraft are of course
synonyms in this section), while Alice locks her spins to her
laboratory on Earth. After Bob's classical computers crashed, far
away from home, the state of the $2N$ spins is still given by
$\ket{\Phi}$ but Alice's and Bob's parts now refer to their
respective SFs. Relative to Bob's SF this state can be written as
\begin{equation}
\ket{\Phi(g)}\equiv U_A(g)\otimes \id_B \ket{\Phi}, \label{Phi(g)}
\end{equation}
where the subscripts $A$ and $B$ refer to Alice's and Bob's
Hilbert spaces respectively and  $g$ stands for the three Euler
angles of the $SO(3)$ rotation that takes Bob's SF into Alice's.
With no other resource available, Alice sends her $N$ spins to
Bob, with the hope that he will retrieve from them the information
he  needs. To do so, he is allowed to perform generalized
collective measurements on both his own spins and Alice's (now in
his possession), namely, on the state~\eqref{Phi(g)}.

In Ref.~\cite{ajv} it was demonstrated that the maximally
entangled state
\begin{equation}\label{state-phi-twit}
    \ket{\tilde{\Phi}}=\sum_j a_j\ket{\tilde{\Phi}^j},
\end{equation}
with
\begin{equation}
\ket{\tilde{\Phi}^{j}}={1\over\sqrt{d_j n_{j}}}\sum_{
m\al}\ket{jm\al}_A\ket{jm\al}_B,
\end{equation}
is the optimal encoding state of an $SU(2)\simeq SO(3)$ operation (notice that
this result does not preclude the existence of other optimal
states).
As in previous sections,  the coefficients $a_j$ have to be
properly chosen to maximize $\aver{\Chi1}$.
It is easy to prove that with this setup the equivalent
representations do not play any role\footnote{One just has to
realize  that any state in the space spanned by the set
$\{U_A(g)\otimes\id_B \ket{\tilde{\Phi}^j}\}$ is orthogonal to a
space spanned by $(n_j-1)\times d_j$ mutually orthogonal states.
Hence, $\ket{\tilde{\Phi}^{j}}$ effectively lives in only one of
the equivalent spin-$j$ representations and can be chosen as in
Eq.~\eqref{state Phi}.}, so, without loss of generality, the
optimal messenger state can be chosen as
\begin{equation}
\ket{\Phi}=\sum_j a_j\ket{\Phi^j}=\sum_{j}{ a_j\over\sqrt{d_j}}
\sum_{m=-j}^j\ket{jm}_A\ket{jm}_B . \label{state Phi}
\end{equation}

Bob's optimal measurement is defined by the (once again) covariant POVM
\begin{equation}
O(g)= \;U_A(g)\otimes \id_B \ket{\Psi}\bra{\Psi}\;U^\dagger_A(g)\otimes\id_B ,
\label{Og-opt}
\end{equation}
where $\ket{\Psi}$ can be taken to be the maximally entangled
state
\begin{equation}\label{Psi-opt}
\ket{\Psi}=\sum_{jm}\sqrt{d_j}\ket{jm}_A\ket{jm}_B.
\end{equation}
In analogy with Eq.~\eqref{aver-t}, $\aver{\Chi1}$ is given by
\begin{equation}
    \aver{\Chi1}=\int dg\,\bra{\Phi}O(g)\ket{\Phi}\Chi1(g)=
    \int dg\,\Chi1(g) \Big|\sum_j
     a_j\Chi1(g)\Big|^2 .
    \label{aver-t-opt}
\end{equation}
The group integral can be easily performed by recalling the
Clebsch-Gordan series $\Chi j(g)\Chi l(g)=\sum_{k=|j-l|}^{j+l}
\Chi{k}(g)$ of $SO(3)$ and the orthogonality of the
characters~\cite{group}, namely, $\int dg \Chi j(g)\Chi
l(g)=\delta_{jl}$. Again, the result can be conveniently written
as in~(\ref{qform}), where now $\mathsf{M}$ is the tri-diagonal
matrix
\begin{equation}\label{matrix-opt}
\mathsf{M}=\pmatrix{
 0  &\zzz \zzz1             &         &   &  &  &\xxx\cr
1&\zzz\zzz 0                               & \zzz \zzz 1     &   & & \mbox{\LARGE0} &\xxx\cr
                   &\zzz\zzz   \raisebox{1.3ex}[0ex][0ex]{1}&\zzz\zzz\raisebox{1.3ex}[0ex][0ex]{0}&\zzz\zzz  \ddots &\zzz \ddots & &\xxx\cr
                   &   &                             &\zzz \ddots  &       &\zzz \zzz1&\xxx\cr
 \hspace{.5cm} \raisebox{2.0ex}[1.5ex][0ex] {\LARGE0}\hspace{-.5cm} & & & &\zzz\zzz  1  &\zzz\zzz 0 &\zzz \zzz1\xxx\cr
                   &   &   &  &  & \zzz\zzz   1  &   \zzz  \zzz    \zeta\xxx
                   } \ ,
\end{equation}
with $\zeta=-1$  ($\zeta=0$) for $N$ even  (odd).

Proceeding as in the previous section, we obtain the maximum value
of $\aver{\Chi1}$ by computing the maximum eigenvalue of
$\mathsf{M}$. The characteristic polynomials $P_n(\lambda)=\det
(\mathsf{M}+2\lambda\id)$, where $n$ is the dimension of $\mathsf
M$ (one has $n=N/2+1$ for $N$ even and $n=N/2+1/2$ for $N$ odd)
again satisfy the recursion relation of the Tchebychev
polynomials, and the solution for the initial conditions derived
from~(\ref{matrix-opt}) is $P_n(\la)=U_n(\la)+\zeta U_{n-1}(\la)$.
The largest eigenvalue of $\mathsf{M}$ is $2\cos[2\pi/(N+3)]$,
hence,
\begin{equation}
\langle\carac1\rangle_{\rm max}=1+2\cos{2\pi\over N+3} ,
\label{chi1_max}
\end{equation}
and one can also verify that the corresponding eigenvector is
\begin{equation}
a_j={2\over\sqrt{N+3}}\sin{(2j+1)\pi\over N+3}  . \label{a_j}
\end{equation}
Remarkably this is an {\em exact} closed solution valid for {\em any} $N$.
For large $N$ this expression reads
\begin{equation}
\aver{\Chi1}_{\rm max}=3 -\frac{4\pi^2}{N^2} +\dots . \label{chi1-asympt}
\end{equation}
Eq.~\eqref{chi1_max} gives the minimum error one can ever attain
when aligning SF through a quantum channel and proves that the
protocol of the previous section, which does not use shared
entanglement between Alice and Bob and requires half the number of
spins, is also optimal {\em asymptotically}.

\section{Conclusions}

We have reviewed  the alignment of spatial frames using quantum
states; a problem that interested Asher in his later years. We
have obtained the optimal protocols assuming different setups. In
all the cases, perfect alignment is possible in the asymptotic
limit. We have also shown that entanglement ---either of internal
degrees of freedom or shared entanglement between Alice and Bob---
dramatically improves the efficiency of the communication of
frames.

A comment on the intrinsicallity of the protocols discussed in
this paper  is in order. It is clear that this communication
problem requires that the encoding of the information about the
frame (``unspeakable" information in Asher's terminology) on the
messenger state (as well as the decoding via measurements) has to
be accomplished through spatial rotations, since no reference is
assumed to be shared by sender and recipient. This is the reason
why we consider spin or angular momentum. By the same reason, the
whole procedure needs be covariant. The precise implementation of
the communication protocols is, of course, well beyond the scope
of this paper. Here we have just presented theoretical lower
bounds on the communication error.

To end, we would like to point out that the problem of aligning
spatial frames can also be translated into the reverse engineering
problem of estimating an unknown $SU(2)$ operation on qubits. We
refer the interested reader to Refs.~\cite{ajv}
and~\cite{bbm-repeated} for details.

\section*{Acknowledgements}
We are grateful to M.~Baig for sharing many contributions with us
along all these years. We acknowledge financial support from
Spa\-nish Ministry of Science and Technology project
BFM2002-02588, CIRIT project SGR-00185, and QUPRODIS working group
EEC contract IST-2001-38877.

\newcommand{\PRL}[3]{Phys.~Rev.~Lett.~\textbf{#1}, #2~(#3)}
\newcommand{\PRA}[3]{Phys.~Rev. A~\textbf{#1}, #2~(#3)}
\newcommand{\JPA}[3]{J.~Phys. A~\textbf{#1}, #2~(#3)}
\newcommand{\PLA}[3]{Phys.~Lett. A~\textbf{#1}, #2~(#3)}
\newcommand{\JOB}[3]{J.~Opt. B~\textbf{#1}, #2~(#3)}
\newcommand{\JMP}[3]{J.~Math.~Phys.~\textbf{#1}, #2~(#3)}
\newcommand{\JMO}[3]{J.~Mod.~Opt.~\textbf{#1}, #2~(#3)}

\end{document}